\documentclass[a4paper,notitlepage,11pt]{article}
\usepackage{a4,layout}
\usepackage[T1]{fontenc}
\usepackage{amsmath,amsfonts,amssymb}
\usepackage{ae,aecompl}
\usepackage[english]{babel}
\usepackage{longtable}
\usepackage[dvips]{graphics}
\usepackage{graphicx}
\usepackage{verbatim}
\usepackage{afterpage}
\usepackage{float}
\usepackage{epsfig}
\usepackage{dsfont}
\usepackage{fancybox}
\usepackage{fancyhdr}
\usepackage{color}
\usepackage{amsmath}
\usepackage{amsfonts}
\usepackage{amssymb}

\newtheorem{theorem}{Theorem}[section]
\newtheorem{lemma}[theorem]{Lemma}
\newtheorem{corollary}[theorem]{Corollary}
\newtheorem{proposition}[theorem]{Proposition}

\fancypagestyle{mystyle}{ 
\lhead{\textit{Analysis of the sensitivity to discrete dividends : A new approach for pricing vanillas}}  
\chead{} 
\rhead{} 
\lfoot{Gocsei-Sahel} 
\cfoot{\bfseries \thepage} 
\rfoot{\bfseries} 
} 
\definecolor{gris25}{gray}{0.75} 
\DeclareMathOperator*{\Call}{Call}
\DeclareMathOperator*{\Proxy}{Proxy}

\DeclareMathOperator*{\cbs}{Call^{BS}}
\DeclareMathOperator*{\pbs}{Put^{BS}}

\begin{document}

\title{Analysis of the sensitivity to discrete dividends : \\
A new approach for pricing vanillas\footnote{The opinions expressed in this article are those of the authors alone,
and do not reflect the views of Société Générale, its subsidiaries or affiliates.}}
\author{Arnaud Gocsei\footnote{Arnaud Gocsei is a quantitative analyst in the Equity model validation team at Société Générale.
E-mail: arnaud.gocsei@sgcib.com.}, Fouad Sahel\footnote{Fouad Sahel is head of the Equity model validation team at Société Générale. E-mail:
mohammed-fouad.sahel@sgcib.com.}}
\date{5 May 2010}
\maketitle

\pagestyle{mystyle}

\begin{center}
\textbf{Abstract}
\end{center}

The incorporation of a dividend yield in the classical option pricing model
of Black-Scholes results in a minor modification of the Black-Scholes
formula, since the lognormal dynamic of the underlying asset is preserved.
However, market makers prefer to work with cash dividends with fixed value
instead of a dividend yield. Since there is no closed-form solution for the
price of a European Call in this case, many methods have been proposed in
the literature to approximate it. Here, we present a new approach. We derive
an exact analytic formula for the sensitivity to dividends of an European
option. We use this result to elaborate a proxy which possesses the same
Taylor expansion around 0 with respect to the dividends as the exact
price. The obtained approximation is very fast to compute (the same
complexity than the usual Black-Scholes formula) and numerical tests show
the extreme accuracy of the method for all practical cases.

\newpage

\tableofcontents

\newpage

\section{Introduction}

\label{intro}

In the classical Black-Scholes framework, we can find in the literature
three main ways of inserting cash dividends into the model \footnote{%
We take here the terminology used in~\cite{vellekoop}.} :

\begin{enumerate}
\item \textbf{Escrowed model.} Assume that the asset price minus the present
value of all dividends to be paid before the maturity of the option follows
a Geometric Brownian Motion.

\item \textbf{Forward model.} Assume that the asset price plus the forward
value of all dividends from past dividend dates to today, follows a
Geometric Brownian Motion.

\item \textbf{Piecewise lognormal model.} Assume that the asset price shows
a jump downward at each dividend date (equal to the cash dividend payment at
that date) and follows a Geometric Brownian Motion between those dates.
\end{enumerate}

Although the first two models lead to a closed-form solution, they are not
satisfactory. Indeed, the option price obtained in these models is
not continuous at dividend dates. Moreover, if one considers two options
with different maturities $T_{1}<T_{2}$, the first two models lead to
different asset price process dynamics for $t\leq T_{1}$, since the
dividends paid between $T_{1}$ and $T_{2}$ are taken into account in one
case but not in the other.

Therefore, it is the piecewise lognormal model which is prefered from a
theoritical point of view. This paper is dedicated to find a robust pricing
proxy for this model. We consider an underlying following a Black-Scholes
dynamic between dividend detachement dates and paying cash dividends at
discrete times $0<T_{1}<\ldots <T_{n}<T$, i.e. :

\begin{itemize}
\item for $T_i\leq t<T_{i+1}$ : 
\begin{equation*}
dS_{t} = r S_{t}dt + \sigma S_{t} dW_{t},
\end{equation*}

\item at time $T_i$ : 
\begin{equation*}
S_{T_i^+} = S_{T_i^-} - D_i (S_{T_i^-}),
\end{equation*}
\end{itemize}

where $r$ is the interest rate, assumed constant, $W$ is a standard Brownian
motion and $D_{i}$ is the dividend policy defined by : 
\begin{equation*}
D_{i}(S)=\left\{ 
\begin{array}{lll}
C_{i} & \text{if} & S>C_{i}, \\ 
S & \text{if} & S\leq C_{i},%
\end{array}%
\right.
\end{equation*}%
The cash amounts $C_{1},\ldots ,C_{n}$ are \emph{known} at the initial date
0 and each $C_i$ represents the dividend cash amount eventually paid at time 
$T_{i}$.

The dividend policy $D_{i}$ is a \emph{liquidator} policy as the stock price
is absorbed at zero at time $T_{i}$ if $S_{T_{i}}<C_{i}$. Consequently, the
stock price remains positive. Note that as a practical matter, for most
applications, the definition of $D_{i}(S)$ when $S\leq C_{i}$ has negligible
financial effects\footnote{%
This becomes less true when considering large maturities and dividends.}, as
the probability that a stock price drops below a declared dividend at a
fixed time is typically small. It just ensures the positivity of the price.

In this paper, we are interested in computing the fair price of the European
Call $\Call(S_{0},K)$ with strike $K$ and maturity $T$. Since there is no 
closed-formula, one should recover the price via PDE methods using a
finite difference scheme, with boundary conditions at each $T_{i}$ ensuring
the continuity of the price of the Call. This procedure can be
time-consuming if one considers a maturity $T=20$ years and an underlying
paying as much as one dividend a week. Therefore, when computation speed is
at stake, one would prefer a fast and accurate proxy for the price.

We review in the following section three of the existing methods in the
literature and discuss their limitations.

\section{Existing Methods}

\begin{enumerate}
\item \textbf{Method of moments matching.} We approximate the stock price
process $S$ by a process $\tilde{S}$ with a shifted log-normal dynamic under
the risk-neutral pricing measure : 
\begin{equation*}
\tilde{S}_{t}=\lambda +M\exp \left( -\frac{1}{2}\sigma ^{^{\prime
}2}t+\sigma ^{^{\prime }}W_{t}\right) .
\end{equation*}%
The three parameters $\lambda ,M$ and $\sigma ^{^{\prime }}$ are calibrated
so that the first three moments of $\tilde{S}_{T}$ match the first three
moments of $S_{T}$. This method reduces to the pricing of a European Call on
a modified underlying $\tilde{S}$, which can be done using the usual
Black-Scholes formula. This proxy does not work well if the stock pays
dividends frequently, the maturity is greater than 5 years or the option is
deep in-the-money.

\item In~\cite{bos2002}, Bos and Vandermark define a mixture of the Escrowed
and Forward models, using linear pertubations of first order. They derive a
proxy resulting in spot/strike adjustment : 
\begin{equation*}
\Call(S_{0},K)\approx \cbs(S^{\ast },K^{\ast }),
\end{equation*}%
where $\cbs$ is the usual Black-Scholes function and: 
\begin{align}
S^{\ast }=& S_{0}-\sum_{i=1}^{n}\left( 1-\frac{T_{i}}{T}\right)
C_{i}e^{-rT_{i}},  \label{Setoile} \\
K^{\ast }=& K+\sum_{i=1}^{n}\frac{T_{i}}{T}C_{i}e^{r(T-T_{i})}.
\label{Ketoile}
\end{align}%
This proxy works better for at-the-money options and small maturities but
results in serious mis-pricing for in-and out-of-the-money options and large
maturities.

\item In~\cite{bos2003}, Bos, Gairat and Shepeleva derive a more accurate
proxy than the previous one by considering a volatility adjustment : 
\begin{equation*}
\Call(S_{0},K)\approx \cbs(S^{\ast },K,\sigma (S^{\ast },K,T)),
\end{equation*}%
with $S^{\ast }$ given by~(\ref{Setoile}): 
\begin{align*}
\sigma (S^{\ast },K,T)^{2}= & \sigma ^{2}+\sigma \sqrt{\frac{\pi }{2T}}%
\Bigg\{\frac{e^{\frac{a^{2}}{2}}}{S^{\ast }}\sum_{i=1}^{n}C_{i}e^{-rT_{i}}%
\left[ N(a)-N\left( a-\sigma \frac{T_{i}}{\sqrt{T}}\right) \right] \\
+& \frac{e^{\frac{b^{2}}{2}}}{S^{\ast 2}} \sum_{i,j=1}^{n}
C_{i}C_{j}e^{-r(T_{i}+T_{j})}\left[ N(b)-N\left( b-2\sigma \frac{%
min(T_{i},T_{j})}{\sqrt{T}}\right) \right],
\end{align*}%
where $N(x)$ is the normal distribution function and: 
\begin{equation*}
a=\frac{1}{\sigma \sqrt{T}}\left( \log \left( \frac{S^{\ast }}{K}\right)
+(r-\sigma ^{2}/2)T\right) ,\ b=a+\frac{1}{2}\sigma \sqrt{T}.
\end{equation*}
This proxy will be a good benchmark to test the accuracy of our method
presented in the following section.
\end{enumerate}

\section{The method}

\subsection{Motivations and notations}

Consider $\Call(S_0,K)$ as a function of the dividends $C_{1},\ldots ,C_{n}$: 
\begin{equation*}
\Call(S_{0},K)=\Call(C_{1},\ldots ,C_{n}).
\end{equation*}
Although there is no closed-form formula for $\Call(C_1,\ldots,C_n)$, we
prove in annex~\ref{annex1} that we can still compute explicitely its
sensitivities to dividends at the origin. More precisely, we have for all $%
k\in \mathbb{N}$ and $1\leq i_{1},\ldots,i_{k}\leq n$:

\begin{equation}  \label{derivative}
\frac{\partial^k \Call}{\partial C_{i_1}\ldots\partial C_{i_k}}(0)=(-1)^k\frac{%
\partial^k \cbs}{\partial S^k}\left(S_0e^{-\sigma^2\sum_{q=1}^kT_{i_q}},K,T%
\right)e^{-r\sum_{q=1}^kT_{i_q}-\sigma^2\sum_{q=2}^k(q-1)T_{i_q}}.
\end{equation}

We use this result to derive an accurate approximation of $\Call%
(C_{1},\ldots ,C_{n})$. Before explaining our method, we need first to
introduce some notations. For all functions $f$ of $n$ variables $%
x_{1},\ldots ,x_{n}$ and $\forall \alpha \in \mathbb{N}$, we note $T_{\alpha
}f$ the $\alpha^{th}$ order Taylor series at 0 of $f$: 
\begin{equation*}
T_{\alpha }f(x_{1},\ldots ,x_{n}):=\sum_{k=0}^{\alpha }\sum_{i_{1},\ldots
,i_{k}=1}^{n}\frac{x_{i_{1}}\ldots x_{i_{k}}}{i_{1}!\ldots i_{k}!}\frac{%
\partial ^{k}f}{\partial x_{i_{1}}\ldots \partial x_{i_{k}}}(0).
\end{equation*}%
We introduce the space $\mathcal{A}_{\alpha }$ of functions having the same
$\alpha^{th}$ Taylor series at 0 as the function $\Call(C_{1},\ldots
,C_{n})$: 
\begin{equation*}
\mathcal{A}_{\alpha }:=\{f,T_{\alpha }f=T_{\alpha }\Call\}.
\end{equation*}%
The order $\alpha $ quantifies how near is $f(C_{1},\ldots ,C_{n})$ from $%
\Call(C_{1},\ldots ,C_{n})$ when the dividends are small. This precision
increase with $\alpha $.

\begin{figure}[H]
\begin{center}\label{toto}
\includegraphics[width=12cm]{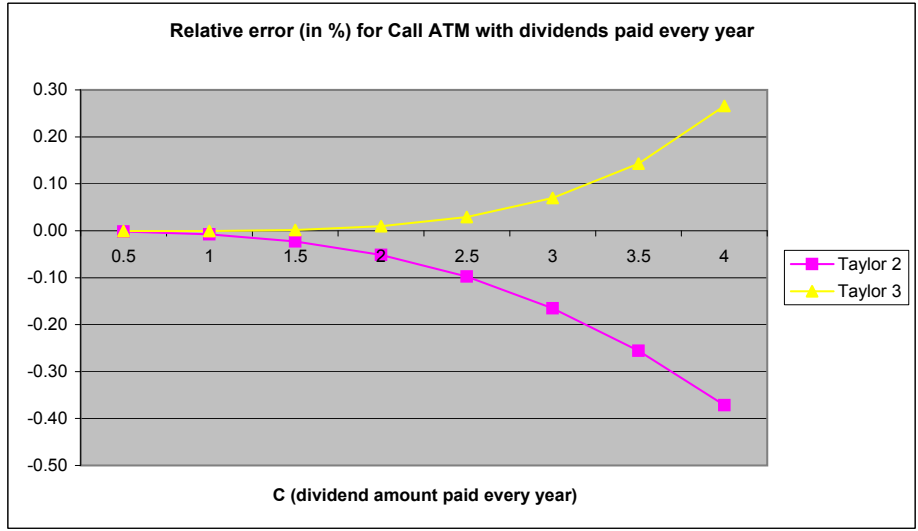}
\end{center}
\caption{Relative Error using the Taylor series for approximation. Numerical parameters: $S_0=100, K=100, r=3\%, \sigma=30\%, T=10y$.}
\end{figure}

Functions $f$ in $\mathcal{A}_{\alpha }$ are naturally good candidates to
approximate $\Call$. However, the difference $\Call(C_{1},\ldots
,C_{n})-f(C_{1},\ldots ,C_{n})$ can be quite big if the dividends are not
small enough. For instance, for $\alpha=2$ and 3, we test the accuracy of the natural choice consisting of taking
\[
f(C_{1},\ldots ,C_{n}):=T_{\alpha}\Call(C_{1},\ldots ,C_{n}).
\]
Figure~\ref{toto} shows the relative error of the price of a European Call when using this approximation. We assume that the stock pays a fixed dividend $C$ every year and we analyse how the relative error varies when we increase $C$. We can see that both the second and third order Taylor series give accurate results when the dividends are small but when the dividends increase, they both lead to serious mis-pricing.

Therefore, the approximations $T_{\alpha }\Call$, $\alpha=2, 3$, are not satisfying. Thus, one need to find for $\alpha \geq 2$, a function $%
\Proxy\in \mathcal{A}_{\alpha }$, different from $T_{\alpha }\Call$, which gives an accurate approximation of $%
\Call$ for all practical values of $C_{1},\ldots ,C_{n}$ (not necessarily
very small). We explain in the next subsection how we determine $\Proxy\in\mathcal{A}_{\alpha}$.

\subsection{Spot/Strike adjustment}

Like Bos and Vandermark in~\cite{bos2002}, we search our function $\Proxy$ under the form: 
\begin{equation}  \label{approxCall}
\Proxy(C_{1},\ldots ,C_{n}):= \cbs(S^{\ast}(C_1,\ldots,C_n),K^{\ast}(C_1,%
\ldots,C_n)),
\end{equation}
with: 
\begin{align}
S^{*}(C_1,\ldots,C_n)=& S_0+\sum_{k=0}^{\alpha}\sum_{i_{1},\ldots,
i_{k}=1}^n a_{i_1,\ldots,i_k}C_{i_1}\ldots C_{i_k},  \label{spotadj} \\
K^{*}(C_1,\ldots,C_n)=& K+\sum_{k=0}^{\alpha}\sum_{i_{1},\ldots, i_{k}=1}^n
b_{i_1,\ldots,i_k}C_{i_1}\ldots C_{i_k}.  \label{strikeadj}
\end{align}
The reason why we perform a spot/strike adjustment is that it allows to
recover the exact price when the dividends are paid spot or at maturity.

The coefficients $a_{i_1,\ldots,i_k}$ and $b_{i_1,\ldots,i_k}$ are
calculated recursively. They are entirely determined by the two following
conditions:

\begin{enumerate}
\item $\frac{\partial^k \Proxy}{\partial C_{i_1}\ldots\partial C_{i_k}}(0)=%
\frac{\partial^k \Call}{\partial C_{i_1}\ldots\partial C_{i_k}}(0), \forall
k\leq \alpha$, 

\item We impose our proxy to satisfy the Call-Put parity\footnote{%
The right term in equation~(\ref{callput}) is not rigorously exact since $%
e^{-rT}E[S_{T}]$ is not equal to $S_{0}-\sum_{i=1}^{n}C_{i}e^{-rT_{i}}$, but
the two quantities are very close.}: 
\begin{align}
\cbs(S^{\ast },K^{\ast })-\pbs(S^{*},K^{*})=&
S_{0}-Ke^{-rT}-\sum_{i=1}^{n}C_{i}e^{-rT_{i}},  \label{callput} \\
i.e.\hspace{3.5cm} S^{\ast }-K^{\ast }e^{-rT}=&
S_{0}-Ke^{-rT}-\sum_{i=1}^{n}C_{i}e^{-rT_{i}}.  \label{eq2}
\end{align}
\end{enumerate}

Let's detail the calculus:

\begin{itemize}
\item \textbf{Computation of $a_i$ and $b_i$:} the equality $\frac{\partial %
\Proxy}{\partial C_i}(0)=\frac{\partial \Call}{\partial C_i}(0)$ reads: 
\begin{equation}
N(d_{1})a_i-e^{-rT}N(d_{2})b_i=-e^{-rT_{i}}N(d(T_{i})),  \label{Eq1}
\end{equation}
where : 
\begin{align*}
d_{1}=& \frac{1}{\sigma \sqrt{T}}(\ln (S_{0}/K)+(r+\sigma ^{2}/2)T), \\
d_{2}=& d_{1}-\sigma \sqrt{T}, \\
d(t)=& d_{1}-\frac{\sigma }{\sqrt{T}}t,\hspace{0.5cm}0\leq t\leq T.
\end{align*}
The differentiation of~(\ref{eq2}) writes: 
\begin{equation}
a_i-e^{-rT}b_i=-e^{-rT_{i}}.  \label{Eq2}
\end{equation}
Solving the linear system~(\ref{Eq1})-(\ref{Eq2}) gives: 
\begin{align*}
a_i=& -e^{-rT_{i}}\frac{N(d(T_{i}))-N(d_{2})}{N(d_{1})-N(d_{2})}, \\
b_i=& e^{r(T-T_{i})}\frac{N(d_{1})-N(d(T_{i}))}{N(d_{1})-N(d_{2})}.
\end{align*}

\item \textbf{Computation of $a_{i,j}$ and $b_{i,j}$:} the equality $\frac{%
\partial^2 \Proxy}{\partial C_i\partial C_j}(0)=\frac{\partial^2 \Call}{%
\partial C_i\partial C_j}(0)$ and two succesive differentiations in~(\ref%
{eq2}) give the following linear system: 
\begin{align}
N(d_{1})a_{i,j}-e^{-rT}N(d_{2})b_{i,j}=& \beta ,  \label{caca} \\
a_{i,j}-e^{-rT}b_{i,j}=& 0,  \label{pipi}
\end{align}
where: 
\begin{align*}
\beta = \frac{\partial ^{2}\Call}{\partial C_{i}\partial C_{j}}(0)- a_ia_j%
\frac{\partial ^{2}\cbs}{\partial S^{2}}(S_{0},K) - (a_ib_j+a_jb_i)\frac{%
\partial ^{2}\cbs}{\partial S\partial K}(S_{0},K) - b_ib_j\frac{\partial ^{2}%
\cbs}{\partial K^{2}}(S_{0},K).
\end{align*}
After some direct computations, we obtain: 
\begin{align*}
a_{i,j}=& \frac{1}{\gamma} e^{-r(T_i+T_j)}\Bigg[a+b\Big(N(d(T_i))+N(d(T_j))%
\Big)+cN(d(T_i))N(d(T_j))+de^{\sigma^2T_i}N^{^{\prime }}(d(T_i+T_j))\Bigg],
\\
b_{i,j}=& e^{rT}a_{i,j},
\end{align*}
with: 
\begin{align*}
\gamma=& \sigma S\sqrt{T}N^{^{\prime }}(d_1)\Big(N(d_1)-N(d_2)\Big)^3 \\
a=& -\Big(N(d_2)N^{^{\prime }}(d_1)-N(d_1)N^{^{\prime }}(d_2)\Big)^2 \\
b=& \Big(N^{^{\prime }}(d_1)-N^{^{\prime }}(d_2)\Big)\Big(N(d_2)N^{^{\prime
}}(d_1)-N(d_1)N^{^{\prime }}(d_2)\Big) \\
c=& -\Big(N^{^{\prime }}(d_1)-N^{^{\prime }}(d_2)\Big)^2 \\
d=& N^{^{\prime }}(d_1)\Big(N(d_1)-N(d_2)\Big)^2 \\
\end{align*}

\item \textbf{Computation of $a_{i_1,\ldots,i_k}$ and $b_{i_1,\ldots,i_k}$, $%
k\geq 3$:} the previous method can be reproduced recursively. Knowing all
the values $a_{j_1,\ldots,j_m}$ and $b_{j_1,\ldots,j_m}$, $m\leq k-1$, we
obtain $a_{i_1,\ldots,i_k}$ and $b_{i_1,\ldots,i_k}$ by solving a linear
system of the form :

\begin{align*}
\frac{\partial^{k}\cbs}{\partial S^k}(S_{0},K) a_{i_1,\ldots,i_k}+\frac{%
\partial^{k}\cbs}{\partial K^k}(S_{0},K)b_{i_1,\ldots,i_k}=& u , \\
a_{i_1,\ldots,i_k}-e^{-rT}b_{i_1,\ldots,i_k}=& 0.
\end{align*}
\end{itemize}

We have presented a simple and general method to derive a function $\Proxy$
in $\mathcal{A}_{\alpha}$ for any $\alpha\in\mathbb{N}$. As for the order $%
\alpha$ that we choose effectively for our tests, the second order
computation is a good choice for performance and accuracy. Before presenting
the numerical results, we recall some desirable properties of our second
order proxy~(\ref{approxCall}):

\begin{enumerate}
\item fast computation, even when one considers a large number $n$ of
dividends.

\item recovery of exact price when all dividends are paid spot or at
maturity.

\item arbitrage free with the Call-Put parity.

\item guarantee of the continuity of the Call price at dividend detachement
dates.

\item accuracy for all practical configurations, even for the extreme cases
(deep in-the-money-option with large maturity and high frequency of
dividends) for which the already existing methods of the financial
literature might lead to serious mis-pricing.
\end{enumerate}

\section{Numerical tests}

\subsection{Test on an underlying paying dividends with low frequency}

We test the accuracy of our proxy on a stock with the following parameters: $%
S_0=100$, $r=3\%$, $\sigma=30\%$. We suppose that the stock pays a dividend
of 3 in the middle of every year. We compute the Call price with strike $%
K\in\{50,75,100,125,150,175,200\}$ and maturity $T\in\{5,10,15,20\}$ using
four methods:

\begin{enumerate}
\item the finite difference method,

\item the method of moments matching.

\item the spot/vol adjustment of Bos, Gairat and Shepeleva\cite{bos2003},

\item our proxy with spot/strike adjustment given by~(\ref{spotadj})-(\ref%
{strikeadj}),
\end{enumerate}

Remember that no approximation is made in the finite difference method. The
results are given in the following tables.

\newpage

\begin{center}
\begin{tabular}{|c|ccccccc|}
\multicolumn{8}{c}{\textbf{Maturity=5 years}} \\ 
\multicolumn{8}{c}{} \\ 
\multicolumn{8}{c}{} \\ \hline
$K/S_0$ & 0.5 & 0.75 & 1 & 1.25 & 1.50 & 1.75 & 2 \\ \hline
\textbf{Price:} &  &  &  &  &  &  &  \\ 
FD (exact price) & 47.14 & 33.85 & 24.42 & 17.79 & 13.12 & 9.79 & 7.39 \\ 
Method of moments & 47.17 & 33.87 & 24.42 & 17.78 & 13.10 & 9.77 & 7.38 \\ 
Proxy BGS & 47.11 & 33.84 & 24.42 & 17.80 & 13.13 & 9.81 & 7.41 \\ 
Proxy GS & 47.14 & 33.85 & 24.42 & 17.79 & 13.12 & 9.79 & 7.39 \\ 
\hline\hline
\textbf{Relative error (in\%):} &  &  &  &  &  &  &  \\ 
Method of moments & 0.07 & 0.06 & 0.01 & -0.05 & -0.11 & -0.16 & -0.20 \\ 
Proxy BGS & -0.05 & -0.05 & -0.02 & 0.03 & 0.10 & 0.17 & 0.24 \\ 
Proxy GS & 0.00 & 0.00 & 0.00 & 0.00 & 0.00 & 0.00 & 0.00 \\ \hline
\end{tabular}

\bigskip \bigskip

\begin{figure}[H]
\begin{center}
\includegraphics[width=12cm]{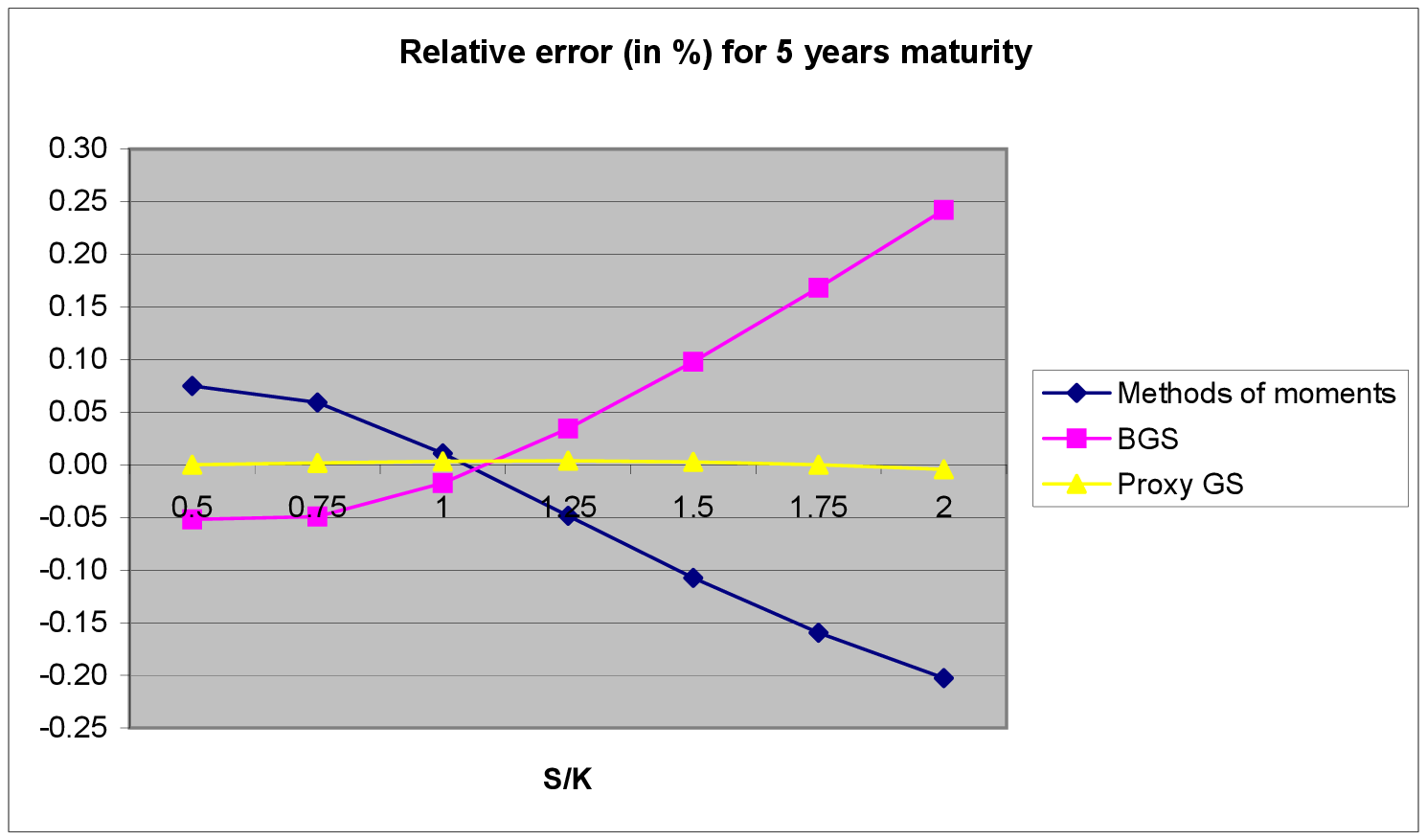}
\end{center}
\end{figure}

\newpage

\begin{tabular}{|c|ccccccc|}
\multicolumn{8}{c}{\textbf{Maturity=10 years}} \\ 
\multicolumn{8}{c}{} \\ 
\multicolumn{8}{c}{} \\ \hline
$K/S_0$ & 0.5 & 0.75 & 1 & 1.25 & 1.50 & 1.75 & 2 \\ \hline
\textbf{Price:} &  &  &  &  &  &  &  \\ 
FD (exact price) & 46.85 & 38.21 & 31.66 & 26.58 & 22.56 & 19.34 & 16.71 \\ 
Method of moments & 47.07 & 38.38 & 31.77 & 26.64 & 22.59 & 19.34 & 16.69 \\ 
Proxy BGS & 46.65 & 38.08 & 31.59 & 26.57 & 22.60 & 19.41 & 16.81 \\ 
Proxy GS & 46.85 & 38.21 & 31.66 & 26.58 & 22.56 & 19.34 & 16.71 \\ 
\hline\hline
\textbf{Relative error (in\%):} &  &  &  &  &  &  &  \\ 
Method of moments & 0.49 & 0.45 & 0.35 & 0.23 & 0.11 & -0.01 & -0.13 \\ 
Proxy BGS & -0.43 & -0.35 & -0.21 & -0.04 & 0.15 & 0.35 & 0.55 \\ 
Proxy GS & 0.00 & -0.01 & -0.01 & -0.01 & -0.01 & -0.01 & -0.02 \\ \hline
\end{tabular}

\bigskip \bigskip

\begin{figure}[H]
\begin{center}
\includegraphics[width=12cm]{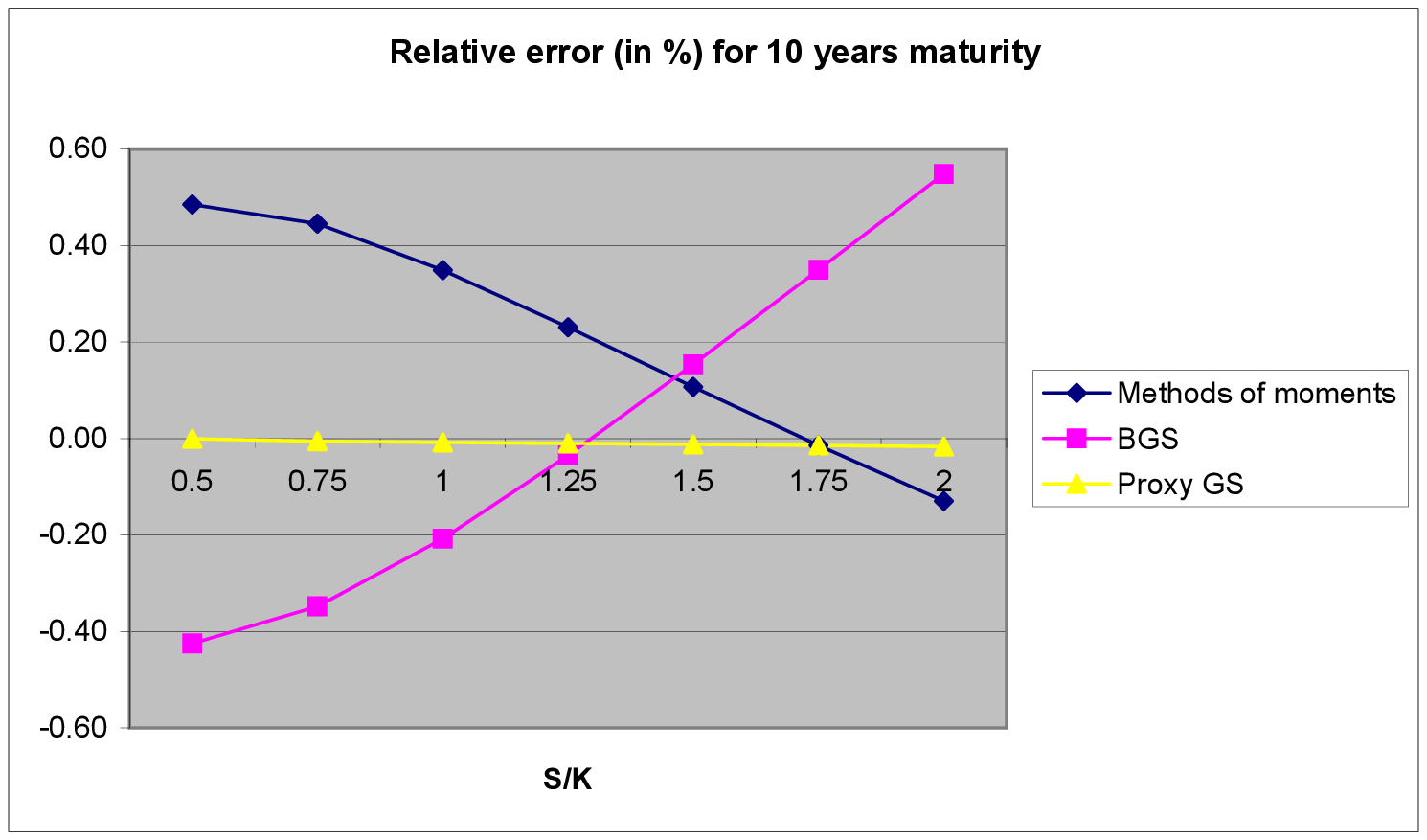}
\end{center}
\end{figure}

\newpage

\begin{tabular}{|c|ccccccc|}
\multicolumn{8}{c}{\textbf{Maturity=15 years}} \\ 
\multicolumn{8}{c}{} \\ 
\multicolumn{8}{c}{} \\ \hline
$K/S_0$ & 0.5 & 0.75 & 1 & 1.25 & 1.50 & 1.75 & 2 \\ \hline
\textbf{Price:} &  &  &  &  &  &  &  \\ 
FD (exact price) & 46.47 & 40.48 & 35.73 & 31.85 & 28.63 & 25.91 & 23.59 \\ 
Method of moments & 47.12 & 41.04 & 36.18 & 32.21 & 28.91 & 26.13 & 23.75 \\ 
Proxy BGS & 45.79 & 40.01 & 35.43 & 31.70 & 28.60 & 25.98 & 23.74 \\ 
Proxy GS & 46.49 & 40.49 & 35.73 & 31.85 & 28.63 & 25.91 & 23.59 \\ 
\hline\hline
\textbf{Relative error (in\%):} &  &  &  &  &  &  &  \\ 
Method of moments & 1.41 & 1.38 & 1.27 & 1.13 & 0.98 & 0.82 & 0.66 \\ 
Proxy BGS & -1.47 & -1.19 & -0.85 & -0.49 & -0.12 & 0.24 & 0.60 \\ 
Proxy GS & 0.02 & -0.01 & -0.02 & -0.03 & -0.03 & -0.04 & -0.04 \\ \hline
\end{tabular}

\bigskip \bigskip

\begin{figure}[H]
\begin{center}
\includegraphics[width=12cm]{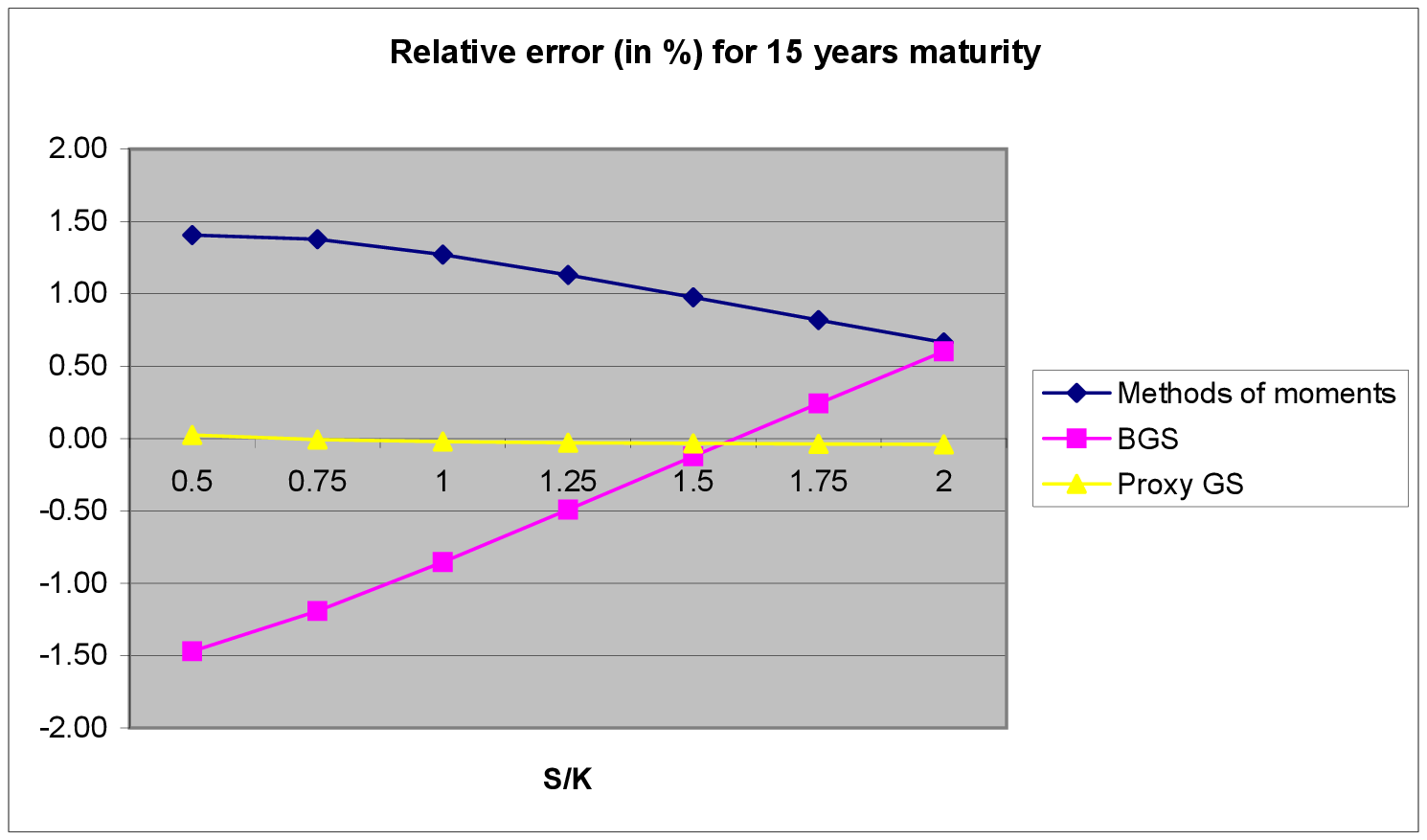}
\end{center}
\end{figure}

\newpage

\begin{tabular}{|c|ccccccc|}
\multicolumn{8}{c}{\textbf{Maturity=20 years}} \\ 
\multicolumn{8}{c}{} \\ 
\multicolumn{8}{c}{} \\ \hline
$K/S_0$ & 0.5 & 0.75 & 1 & 1.25 & 1.50 & 1.75 & 2 \\ \hline
\textbf{Price:} &  &  &  &  &  &  &  \\ 
FD (exact price) & 46.02 & 41.74 & 38.22 & 35.26 & 32.72 & 30.51 & 28.57 \\ 
Method of moments & 47.35 & 42.95 & 39.30 & 36.21 & 33.55 & 31.24 & 29.20 \\ 
Proxy BGS & 44.33 & 40.47 & 37.30 & 34.63 & 32.33 & 30.32 & 28.55 \\ 
Proxy GS & 46.10 & 41.76 & 38.23 & 35.26 & 32.71 & 30.50 & 28.56 \\ 
\hline\hline
\textbf{Relative error (in\%):} &  &  &  &  &  &  &  \\ 
Method of moments & 2.89 & 2.90 & 2.82 & 2.69 & 2.54 & 2.38 & 2.21 \\ 
Proxy BGS & -3.71 & -3.07 & -2.44 & -1.82 & -1.23 & -0.65 & -0.09 \\ 
Proxy GS & 0.14 & 0.03 & -0.02 & -0.04 & -0.05 & -0.06 & -0.07 \\ \hline
\end{tabular}
\end{center}

\bigskip \bigskip

\begin{figure}[H]
\begin{center}
\includegraphics[width=12cm]{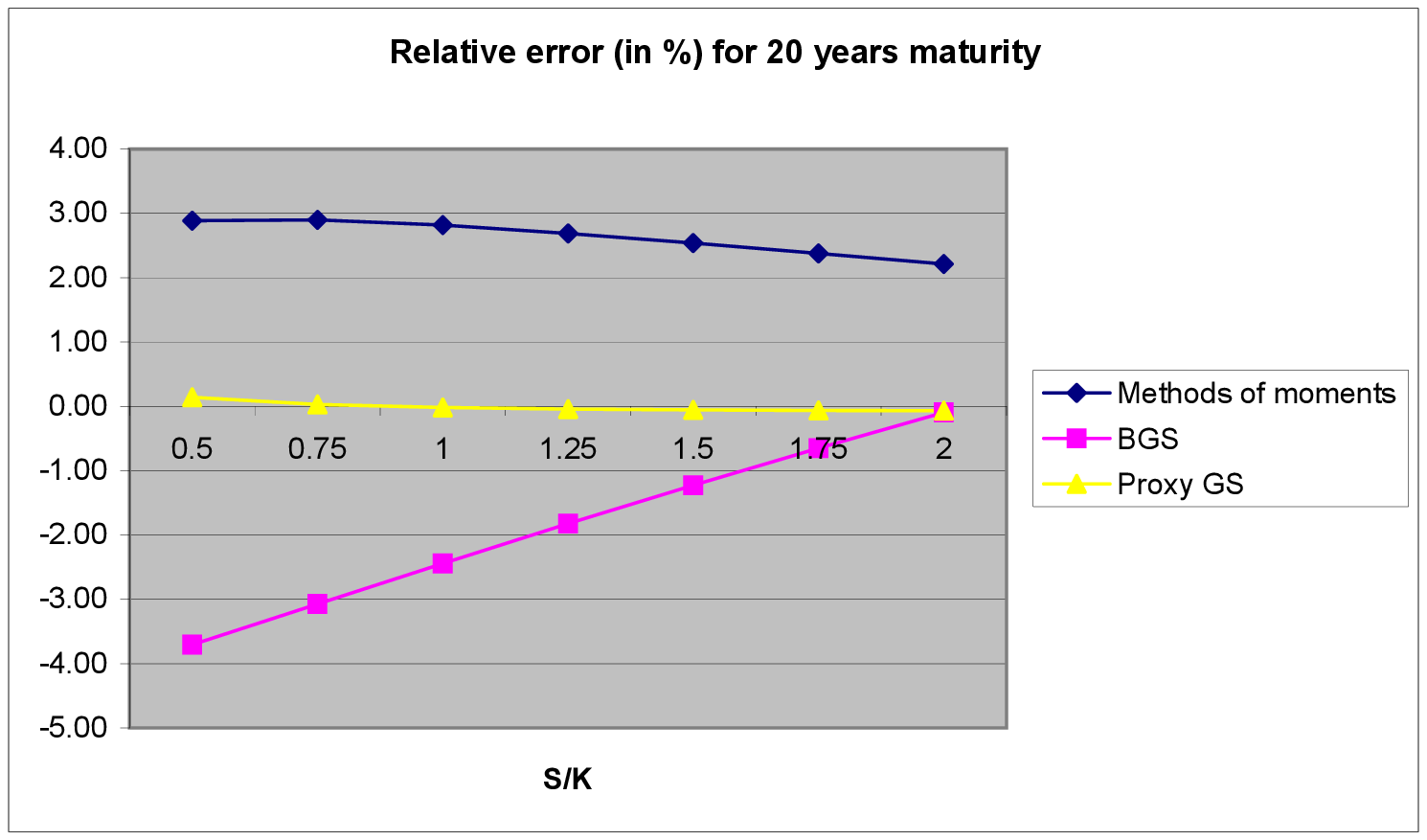}
\end{center}
\end{figure}

\newpage

\subsection{Test on an underlying paying dividends with high frequency}

We now check the accuracy of our proxy on an underlying paying dividends
every week. This situation occurs when considering an index like S\&P 500 or
Eurostox 50. We take the following parameters: $S_0=3000$, $r=3\%$, $%
\sigma=30\%$. We suppose that the stock pays a dividend of 2 every week. We
compute the Call price with strike $K$ such as $K/S_0\in%
\{0.5,0.75,1,1.25,1.5,1.75,2\}$ and maturity $T\in\{5,10,15,20\}$.

\newpage

\begin{center}
\begin{tabular}{|c|ccccccc|}
\multicolumn{8}{c}{\textbf{Maturity=5 years}} \\ 
\multicolumn{8}{c}{} \\ 
\multicolumn{8}{c}{} \\ \hline
$K/S_0$ & 0.5 & 0.75 & 1 & 1.25 & 1.50 & 1.75 & 2 \\ \hline
\textbf{Price:} &  &  &  &  &  &  &  \\ 
FD (exact price) & 1359.87 & 972.67 & 699.65 & 508.71 & 374.45 & 279.07 & 
210.47 \\ 
Method of moments & 1361.05 & 973.29 & 699.70 & 508.39 & 373.97 & 278.54 & 
209.98 \\ 
Proxy BGS & 1358.88 & 972.02 & 699.52 & 508.99 & 375.00 & 279.75 & 211.21 \\ 
Proxy GS & 1359.87 & 972.69 & 699.68 & 508.73 & 374.47 & 279.07 & 210.47 \\ 
\hline\hline
\textbf{Relative error (in\%):} &  &  &  &  &  &  &  \\ 
Method of moments & 0.09 & 0.06 & 0.01 & -0.06 & -0.13 & -0.19 & -0.23 \\ 
Proxy BGS & -0.07 & -0.07 & -0.02 & 0.05 & 0.15 & 0.25 & 0.35 \\ 
Proxy GS & 0.00 & 0.00 & 0.00 & 0.00 & 0.00 & 0.00 & 0.00 \\ \hline
\end{tabular}

\bigskip \bigskip

\begin{figure}[H]
\begin{center}
\includegraphics[width=12cm]{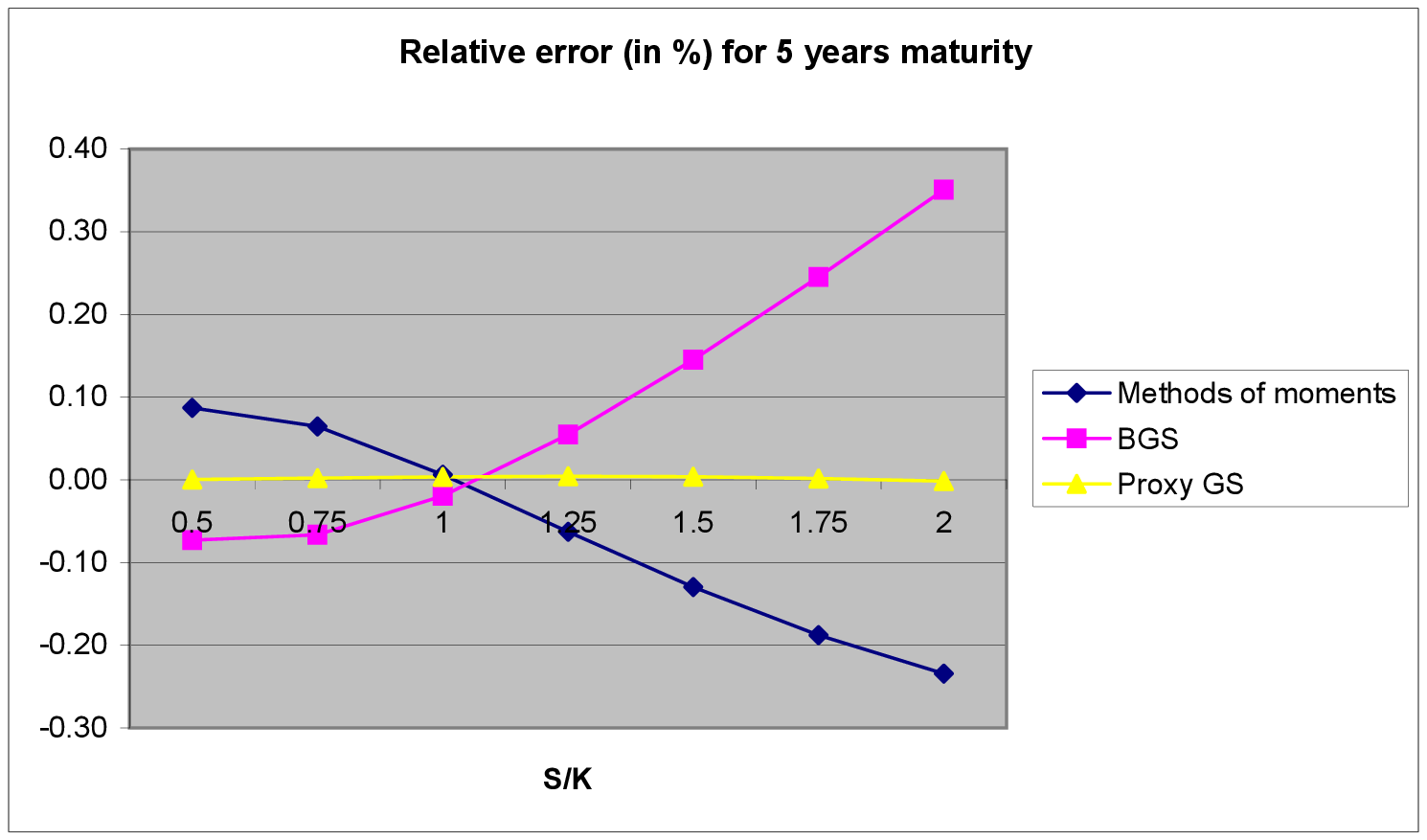}
\end{center}
\end{figure}

\newpage

\begin{tabular}{|c|ccccccc|}
\multicolumn{8}{c}{\textbf{Maturity=10 years}} \\ 
\multicolumn{8}{c}{} \\ 
\multicolumn{8}{c}{} \\ \hline
$K/S_0$ & 0.5 & 0.75 & 1 & 1.25 & 1.50 & 1.75 & 2 \\ \hline
\textbf{Price:} &  &  &  &  &  &  &  \\ 
FD (exact price) & 1319.62 & 1075.07 & 890.03 & 746.82 & 633.81 & 543.14 & 
469.40 \\ 
Method of moments & 1327.05 & 1080.52 & 893.49 & 748.67 & 634.44 & 542.91 & 
468.55 \\ 
Proxy BGS & 1311.37 & 1069.84 & 887.68 & 746.80 & 635.57 & 546.23 & 473.42
\\ 
Proxy GS & 1319.68 & 1075.04 & 889.96 & 746.72 & 633.69 & 543.02 & 469.25 \\ 
\hline\hline
\textbf{Relative error (in\%):} &  &  &  &  &  &  &  \\ 
Method of moments & 0.56 & 0.51 & 0.39 & 0.25 & 0.10 & -0.04 & -0.18 \\ 
Proxy BGS & -0.63 & -0.49 & -0.26 & 0.00 & 0.28 & 0.57 & 0.86 \\ 
Proxy GS & 0.00 & 0.00 & -0.01 & -0.01 & -0.02 & -0.02 & -0.03 \\ \hline
\end{tabular}

\bigskip \bigskip

\begin{figure}[H]
\begin{center}
\includegraphics[width=12cm]{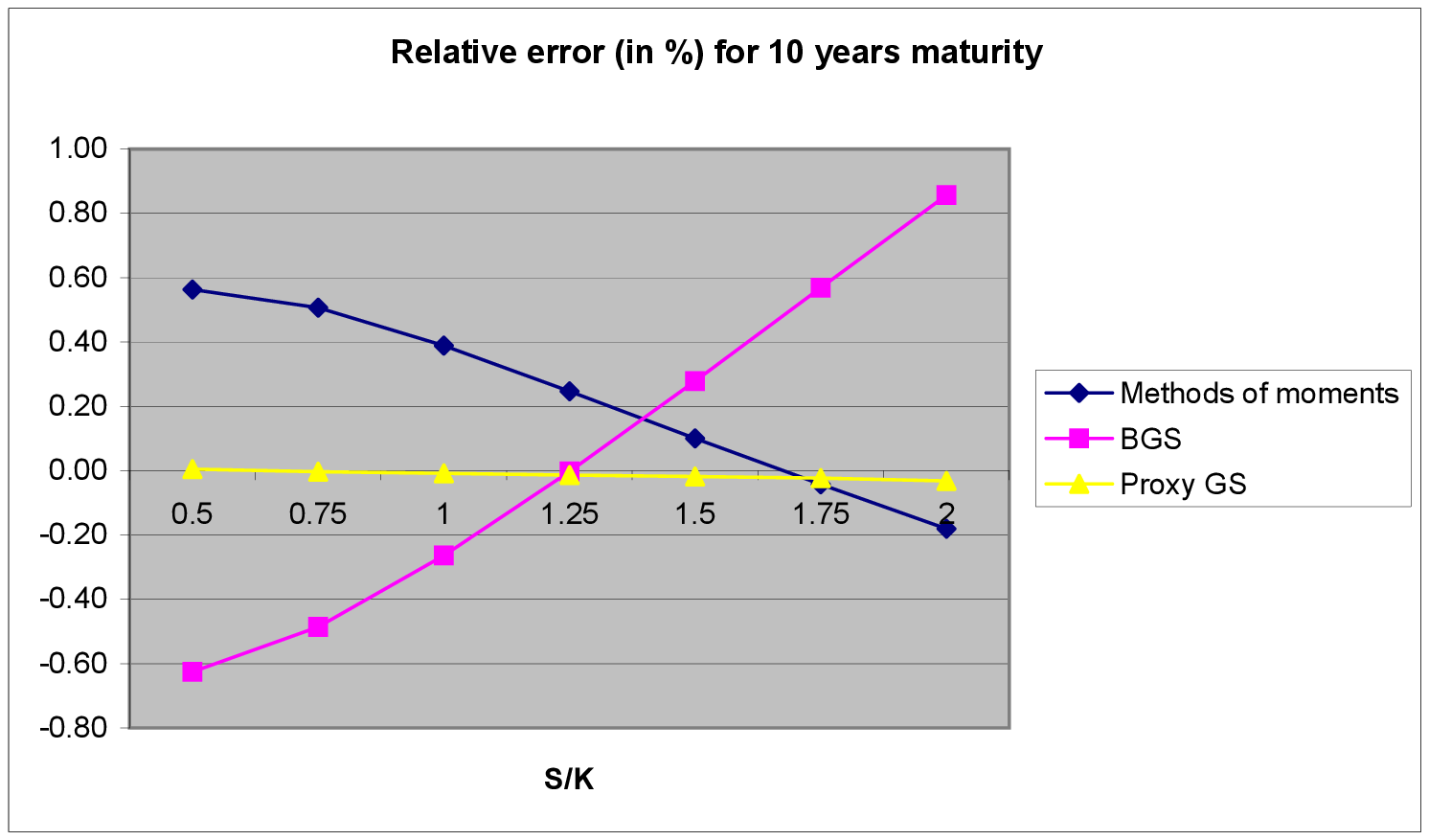}
\end{center}
\end{figure}

\newpage

\begin{tabular}{|c|ccccccc|}
\multicolumn{8}{c}{\textbf{Maturity=15 years}} \\ 
\multicolumn{8}{c}{} \\ 
\multicolumn{8}{c}{} \\ \hline
$K/S_0$ & 0.5 & 0.75 & 1 & 1.25 & 1.50 & 1.75 & 2 \\ \hline
\textbf{Price:} &  &  &  &  &  &  &  \\ 
FD (exact price) & 1287.50 & 1122.13 & 990.74 & 883.63 & 794.68 & 719.49 & 
655.20 \\ 
Method of moments & 1308.65 & 1140.08 & 1005.32 & 895.16 & 803.52 & 726.17 & 
660.09 \\ 
Proxy BGS & 1258.42 & 1102.31 & 978.74 & 877.99 & 794.07 & 723.01 & 662.03
\\ 
Proxy GS & 1288.47 & 1122.33 & 990.66 & 883.42 & 794.31 & 719.10 & 654.79 \\ 
\hline\hline
\textbf{Relative error (in\%):} &  &  &  &  &  &  &  \\ 
Method of moments & 1.64 & 1.60 & 1.47 & 1.31 & 1.11 & 0.93 & 0.75 \\ 
Proxy BGS & -2.26 & -1.77 & -1.21 & -0.64 & -0.08 & 0.49 & 1.04 \\ 
Proxy GS & 0.08 & 0.02 & -0.01 & -0.02 & -0.05 & -0.06 & -0.06 \\ \hline
\end{tabular}

\bigskip \bigskip

\begin{figure}[H]
\begin{center}
\includegraphics[width=12cm]{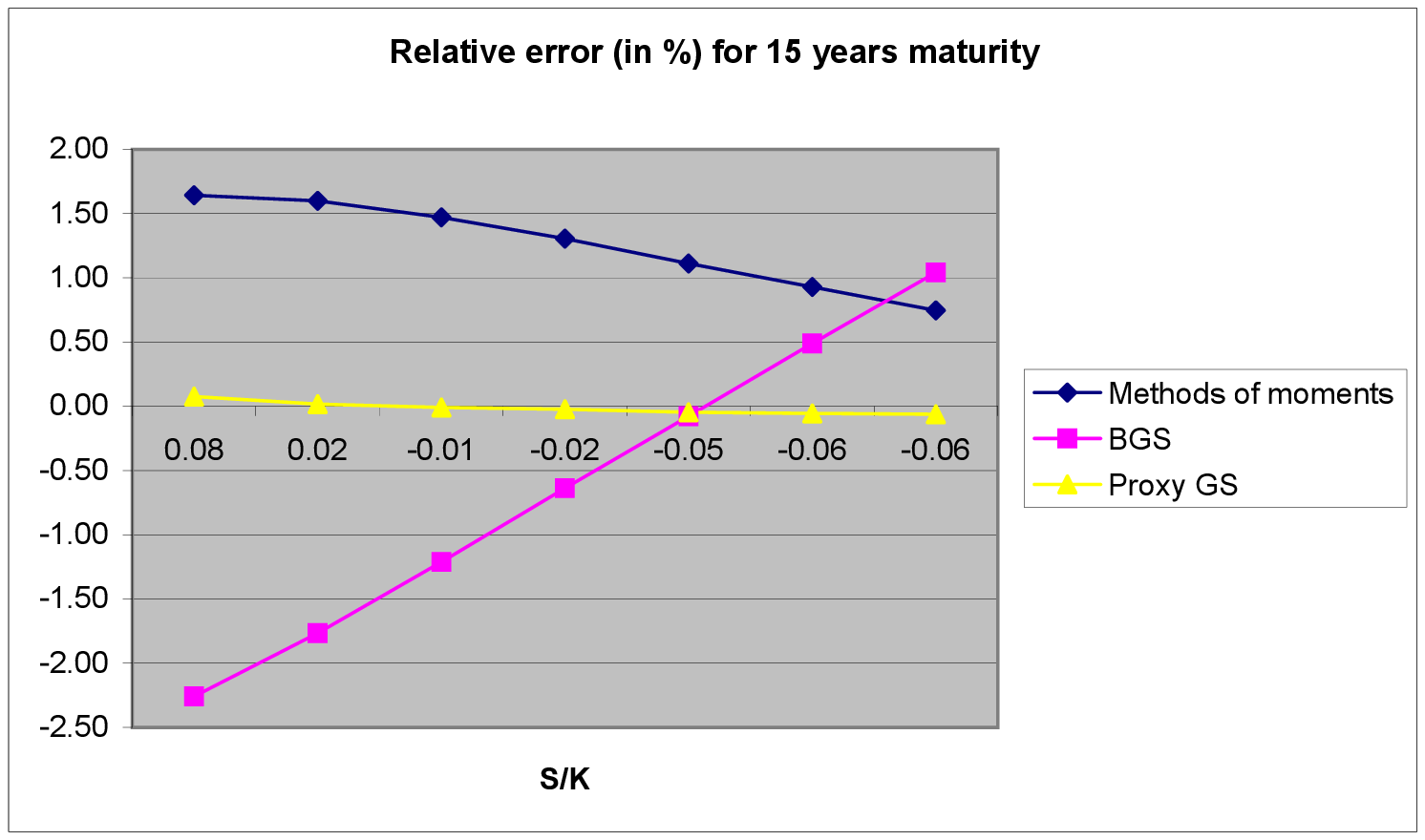}
\end{center}
\end{figure}

\newpage

\begin{tabular}{|c|ccccccc|}
\multicolumn{8}{c}{\textbf{Maturity=20 years}} \\ 
\multicolumn{8}{c}{} \\ 
\multicolumn{8}{c}{} \\ \hline
$K/S_0$ & 0.5 & 0.75 & 1 & 1.25 & 1.50 & 1.75 & 2 \\ \hline
\textbf{Price:} &  &  &  &  &  &  &  \\ 
FD (exact price) & 1260.33 & 1144.53 & 1049.11 & 968.59 & 899.43 & 839.22 & 
786.23 \\ 
Method of moments & 1303.36 & 1183.64 & 1083.92 & 999.27 & 926.36 & 862.79 & 
806.82 \\ 
Proxy BGS & 1184.43 & 1088.21 & 1008.55 & 940.91 & 882.41 & 831.09 & 785.57
\\ 
Proxy GS & 1264.53 & 1145.94 & 1049.44 & 968.43 & 899.04 & 838.71 & 785.66
\\ \hline\hline
\textbf{Relative error (in\%):} &  &  &  &  &  &  &  \\ 
Method of moments & 3.41 & 3.42 & 3.32 & 3.17 & 2.99 & 2.81 & 2.62 \\ 
Proxy BGS & -6.02 & -4.92 & -3.87 & -2.86 & -1.89 & -0.97 & -0.08 \\ 
Proxy GS & 0.33 & 0.12 & 0.03 & -0.02 & -0.04 & -0.06 & -0.07 \\ \hline
\end{tabular}
\end{center}

\bigskip \bigskip

\begin{figure}[H]
\begin{center}
\includegraphics[width=12cm]{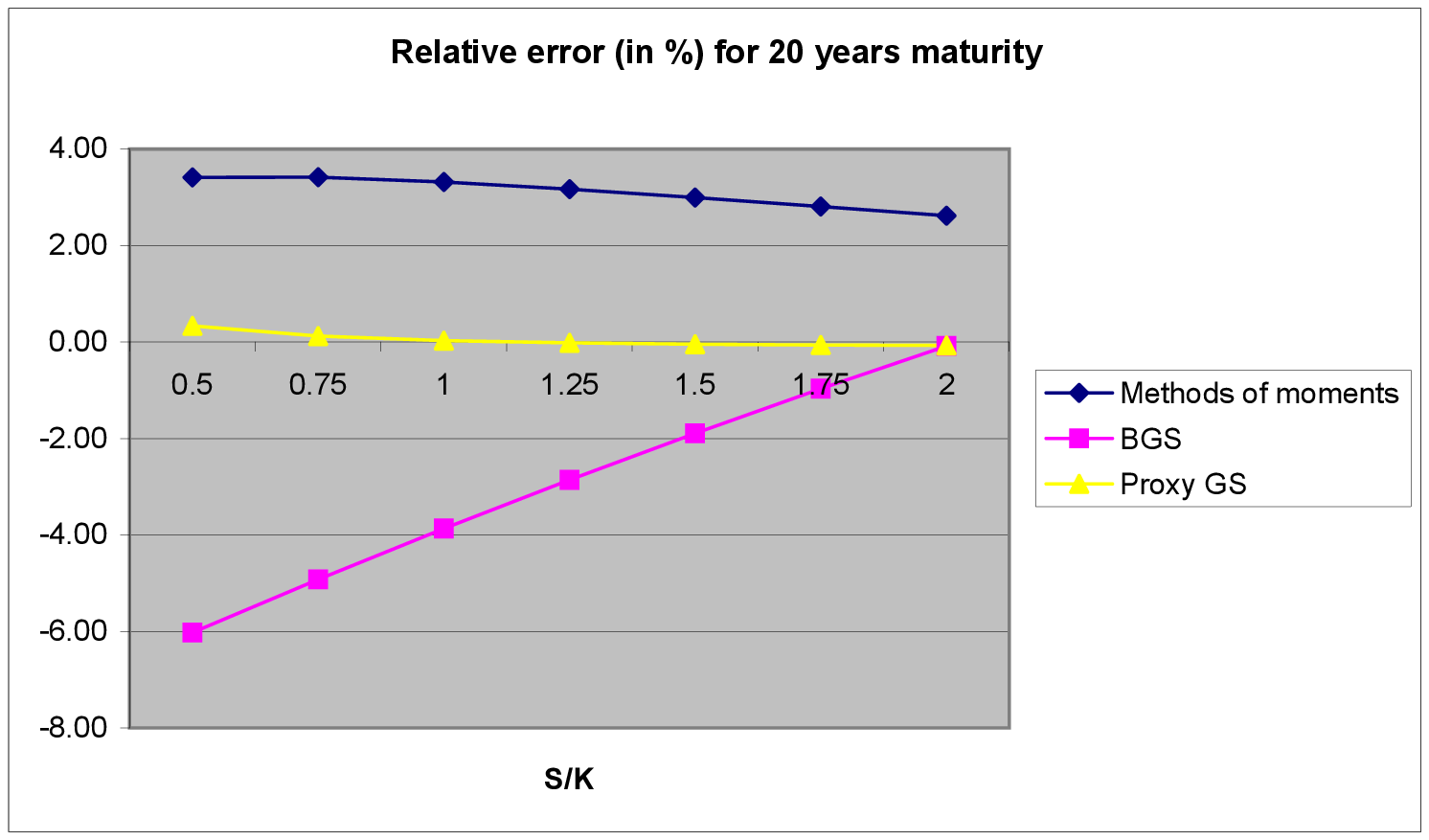}
\end{center}
\end{figure}

\newpage

\section{Conclusion}

We have presented a new approach to deal with cash dividends in equity
option pricing in a piecewise lognormal model for the underlying. Our method
relies on the derivation of an analytic formula for the sensitivity to
dividends of a European option. We obtain a closed-form formula for a
European Call which gives both very accurate results for all practical cases.

\appendix{}

\section{Computation of the dividend sensitivities}

\label{annex1}

Consider a European option of maturity $T$ with payoff $h(S_{T})$, with $S$
the stock price following the piecewise lognormal dynamic presented in the
introduction. We note its fair price at time 0 
\begin{equation*}
\Pi (S_{0},T,C_{1},\ldots ,C_{n})
\end{equation*}%
We denote: 
\begin{equation*}
\Pi ^{BS}(S_{0},T)
\end{equation*}%
the fair price of the option if $S$ does not pay dividends. The partial
derivatives 
\begin{equation*}
\frac{\partial ^{k}\Pi }{\partial C_{i_{1}}\ldots \partial C_{i_{k}}}%
(S_{0},T,0,\ldots ,0)
\end{equation*}%
are related to the usual Black-Scholes greeks by the following formula:

\begin{proposition}
\label{prop} For $k\in\mathbb{N}$ and $1\leq i_1\leq\ldots\leq i_k\leq n$,
we have: 
\begin{align*}
\frac{\partial^k \Pi}{\partial C_{i_1}\ldots\partial C_{i_k}}%
(S_0,T,0,\ldots,0)=(-1)^k\frac{\partial^k \Pi^{BS}}{\partial S^k}%
\left(S_0e^{-\sigma^2\sum_{q=1}^kT_{i_q}},T\right)e^{-r\sum_{q=1}^kT_{i_q}-%
\sigma^2\sum_{q=2}^k(q-1)T_{i_q}}.
\end{align*}
\end{proposition}

Follows a proof of this formula.

\subsection{First step: a recursive formula}

We introduce some notations:

\begin{itemize}
\item We define the natural filtration $(\mathcal{F}_{t})_{t\geq 0}$
associated with the brownian motion $W$. We suppose the filtration right
continuous.

\item We define for all $0\leq t_1\leq t_2$: 
\begin{equation*}
X_{t_1\to t_2}:=e^{(r-\sigma^2/2)(t_2-t_1)+\sigma (W_{t_2}-W_{t_1})},
\end{equation*}

\item We denote $\phi (S_{0},S,t)$ the log-normal density associated with
the variable $S_{0}X_{0\rightarrow t}$

\item We define the functions of $n+1$ variables $(h_i)_{0\leq i \leq n}$
such as: 
\begin{equation*}
h_i(S_{T_i},C_1,\ldots,C_n):=e^{-r(T-T_i)}E[h(S_T)|\mathcal{F}_{T_i}].
\end{equation*}
\end{itemize}

For the sake of simplicity, when there is no confusion, we will simply
denote $h_{i}(S)$ instead of $h_{i}(S,C_{1},\ldots ,C_{n})$. Note that we
have $\Pi (S_{0},T,C_{1},\ldots ,C_{n})=h_{0}(S_{0})$. We can compute the
functions $h_{i}$ recursively beginning with $h_{n}$: 
\begin{equation*}
h_{n}(S)=\Pi ^{BS}(S,T-T_{n}),
\end{equation*}%
and by conditioning, $\forall i\leq n-1$: 
\begin{align}
h_{i}(S)=& e^{-r(T_{i+1}-T_{i})}E[h_{i+1}((SX_{t_{i}\rightarrow
t_{i+1}}-C_{i+1})_{+})|\mathcal{F}_{T_{i}}]  \notag \\
=& e^{-r(T_{i+1}-T_{i})}\int_{C_{i+1}}^{\infty }h_{i+1}(S_{i+1}-C_{i+1})\phi
(S,S_{i+1},T_{i+1}-T_{i})dS_{i+1},  \label{intergal}
\end{align}

Now, we show how these relations allow us to compute recursively the partial
derivatives: 
\begin{equation*}
\frac{\partial ^{k}\Pi }{\partial C_{i_{1}}\ldots \partial C_{i_{k}}}%
(S_{0},0,\ldots ,0),
\end{equation*}%
for $1\leq i\leq n$, $k\in \mathbb{N}^{\ast }$ and $1\leq i_{1}\leq \ldots
\leq i_{k}\leq n$. First, note that a direct application of the theorem of
differentiation under the integral sign in the relation~(\ref{intergal})
proves that the functions $h_{i}$, $0\leq i\leq n$ are infinitely
differentiable. Then, using the markov property of the log-normal densities: 
\begin{equation*}
\int_{0}^{\infty }\phi (S_{i},S_{i+1},t_{i})\phi
(S_{i+1},S_{i+2},t_{i+1})dS_{i+1}=\phi (S_{i},S_{i+2},t_{i}+t_{i+1}),
\end{equation*}%
we obtain: 
\begin{align}
\frac{\partial ^{k}\Pi }{\partial C_{i_{1}}\ldots \partial C_{i_{k}}}%
(S_{0},0,\ldots ,0)=& -e^{-rT_{i_{1}}}E\left[ \frac{\partial ^{k}h_{i_{1}}}{%
\partial S\partial C_{i_{2}}\ldots \partial C_{i_{k}}}(S_{0}X_{0\rightarrow
T_{i_{1}}},0,\ldots ,0)\right] ,  \notag \\
=& -e^{-rT_{i_{1}}}\int_{0}^{\infty }\frac{\partial ^{k}h_{i_{1}}}{\partial
S\partial C_{i_{2}}\ldots \partial C_{i_{k}}}(S_{i_{1}},0,\ldots ,0)\phi
(S_{0},S_{i_{1}},T_{i_{1}})dS_{i_{1}}.  \label{fifi}
\end{align}%
This relation will be very useful for a recursive proof of proposition~\ref%
{prop} since it reduces by one the number of differentiations with respect
to the dividends.

\subsection{Second step: a martingale argument}

The proof of proposition~\ref{prop} relies heavily on this simple but
crucial lemma:

\begin{lemma}
\label{lemma} Consider a process following a Black-Scholes dynamic $%
S_t=S_0e^{(r-\sigma^2/2)t+\sigma W_t}$, $0\leq t\leq T$. Then, for all
integer $k\geq 0$ and for all real number $a\geq 0$, the process: 
\begin{equation*}
Z_t:=\frac{\partial^k \Pi^{BS}}{\partial S^k}\left(S_te^{k\sigma^2(t-a)},T-t%
\right)e^{(k-1)(r+k\sigma^2/2)t}
\end{equation*}
is a martingale.
\end{lemma}

The following corollary is easy to derive.

\begin{corollary}
\label{cor} For $k\in\mathbb{N}^*$, $0\leq t\leq T$ and $a\geq 0$, we have: 
\begin{equation}  \label{bob}
E\left[\frac{\partial^k \Pi^{BS}}{\partial S^k}\left(S_te^{k%
\sigma^2(t-a)},T-t\right)\right]= \frac{\partial^k \Pi^{BS}}{\partial S^k}%
\left(S_0e^{-k\sigma^2a},T\right)e^{-(k-1)(r+k\sigma^2/2)t}
\end{equation}
\end{corollary}

\emph{Proof of lemma~\ref{lemma}:} The drift of $Z_t$ is: 
\begin{align}
&e^{(k-1)(r+k\sigma^2/2)t}\Bigg[(k-1)(r+k\sigma^2/2)\frac{\partial^k \Pi^{BS}%
}{\partial S^k}-\frac{\partial^{k+1}\Pi^{BS}}{\partial t\partial S^k}  \notag
\\
&+(r+k\sigma^2)S_te^{k\sigma^2(t-a)}\frac{\partial^{k+1} \Pi^{BS}}{\partial
S^{k+1}} +\frac{1}{2}\sigma^2S_t^2e^{2k\sigma^2(t-a)}\frac{\partial^{k+2}
\Pi^{BS}}{\partial S^{k+2}}\Bigg],  \label{plop}
\end{align}
where all the derivatives in the last formula are evaluated in $%
(S_te^{k\sigma^2(t-a)},T-t)$. Remember that $\Pi^{BS}$ satisfies the
Black-Scholes PDE: 
\begin{equation*}
-\frac{\partial\Pi^{BS}}{\partial t}+\frac{1}{2}\sigma^2S^2\frac{\partial^2
\Pi^{BS}}{\partial S^2} +rS\frac{\partial \Pi^{BS}}{\partial S}-r\Pi^{BS}=0.
\end{equation*}
Now, differentiate $k$ times this equation with respect to $S$: 
\begin{equation}  \label{BSn}
-\frac{\partial^{k+1}\Pi^{BS}}{\partial t\partial S^k} +\frac{1}{2}%
\sigma^2S^2\frac{\partial^{k+2} \Pi^{BS}}{\partial S^{k+2}} +(r+k\sigma^2)S%
\frac{\partial^{k+1} \Pi^{BS}}{\partial S^{k+1}} +(k-1)(r+k\sigma^2/2)\frac{%
\partial^k \Pi^{BS}}{\partial S^k}=0.
\end{equation}
One immediately checks that the term in bracket in~(\ref{plop}) is equal to
the left term in~(\ref{BSn}) evaluated in $(S_te^{k\sigma^2(t-a)},T-t)$, and
therefore is equal to 0.

\begin{flushright}
$\Box$
\end{flushright}

\subsection{Third step: Proof of proposition~\protect\ref{prop}}

We argue by recurrence on the number $k$ of differentiations with respect to
the dividends: \newline

\begin{itemize}
\item If $k=0$, the proposition is trivially true as it simply says: 
\begin{equation*}
\Pi(S_0,T,0,\ldots,0)=\Pi^{BS}(S,T).
\end{equation*}

\item Now, suppose that the property is true at rank $k$. We want to prove
that it remains true at rank $k+1$. We have by relation~(\ref{fifi}):

\begin{align}  \label{alice}
\frac{\partial^{k+1} \Pi}{\partial C_{i_1} \ldots \partial C_{i_{k+1}}}%
(S_0,0,\ldots,0)=& -e^{-rT_{i_1}}E\left[\frac{\partial^{k+1} h_{i_1}}{%
\partial S\partial C_{i_2} \ldots \partial C_{i_{k+1}}}(S_0X_{0\to
T_{i_1}},0,\ldots,0)\right].
\end{align}
By hypothesis of recurrence, we have:

\begin{align*}
& \frac{\partial ^{k}h_{i_{1}}}{\partial C_{i_{2}}\ldots \partial C_{i_{k+1}}%
}(S,0,\ldots ,0) \\
& =(-1)^k \frac{\partial ^{k}\Pi ^{BS}}{\partial S^{k}}\left( Se^{-\sigma
^{2}\sum_{q=2}^{k+1}(T_{i_{q}}-T_{i_{1}})},T-T_{i_{1}}\right) \times
e^{-r\sum_{q=2}^{k+1}(T_{i_{q}}-T_{i_{1}})-\sigma
^{2}\sum_{q=3}^{k+1}(q-2)(T_{i_{q}}-T_{i_{1}})}, \\
& =(-1)^k \frac{\partial ^{k}\Pi ^{BS}}{\partial S^{k}}\left( Se^{(k+1)\sigma
^{2}(T_{i_{1}}-a )},T-T_{i_{1}}\right) \times e^{\left[ (k+1)r+\frac{1}{2}%
k(k-1)\sigma ^{2}\right] T_{i_{1}}-(k+1)ra -\sigma
^{2}\sum_{q=3}^{k+1}(q-2)T_{i_{q}}},
\end{align*}%
where we set: 
\begin{equation*}
a =\frac{1}{k+1}\sum_{q=1}^{k+1}T_{i_{q}}.
\end{equation*}%
We differentiate with respect to $S$: 
\begin{align*}
& \frac{\partial ^{k+1}h_{i_{1}}}{\partial S\partial C_{i_{2}}\ldots
\partial C_{i_{k+1}}}(S,0,\ldots ,0) \\
& =(-1)^k \frac{\partial ^{k+1}\Pi ^{BS}}{\partial S^{k+1}}\left( Se^{(k+1)\sigma
^{2}(T_{i_{1}}-a )},T-T_{i_{1}}\right) \times e^{(k+1)\left( r+\frac{1}{2}%
k\sigma ^{2}\right) T_{i_{1}}-(k+1)(r+\sigma ^{2})a -\sigma
^{2}\sum_{q=3}^{k+1}(q-2)T_{i_{q}}}.
\end{align*}

Inserting this formula into~(\ref{alice}) and using corollary~\ref{cor}, we
get: 
\begin{align*}
& \frac{\partial ^{k+1}\Pi }{\partial C_{i_{1}}\ldots \partial C_{i_{k+1}}}%
(0,\ldots ,0) \\
=& (-1)^{k+1} \frac{\partial ^{k+1}\Pi ^{BS}}{\partial S^{k+1}}\left( Se^{-(k+1)\sigma
^{2}a },T\right) \exp \left( -(k+1)(r+\sigma ^{2})a -\sigma
^{2}\sum_{q=3}^{k+1}(q-2)T_{i_{q}}\right) , \\
=& (-1)^{k+1} \frac{\partial ^{k+1}\Pi ^{BS}}{\partial S^{k+1}}\left( Se^{-\sigma
^{2}\sum_{q=1}^{k+1}T_{i}},T\right) e^{-r\sum_{q=1}^{k+1}T_{i_{q}}-\sigma
^{2}\sum_{q=2}^{k+1}(q-1)T_{i_{q}}}.
\end{align*}
\end{itemize}

\begin{flushright}
$\Box$
\end{flushright}

\end{document}